\documentclass[twocolumn,superscriptaddress]{revtex4}

\usepackage{amssymb}
\usepackage{amsmath}
\usepackage{multirow}

\usepackage{graphicx}
\usepackage{dcolumn}
\usepackage{bm}
\newfont{\Fr}{eufm10}

\begin{document}

\title{Dependence on supernovae light-curve processing in void models}
\author{Gabriel R. Bengochea}
\email{gabriel@iafe.uba.ar} \affiliation{Instituto de Astronom\'\i
a y F\'\i sica del Espacio (IAFE), UBA-CONICET, CC 67, Suc. 28,
1428 Buenos Aires, Argentina}
\author{Maria E. De Rossi}
\email{derossi@iafe.uba.ar} \affiliation{Instituto de Astronom\'\i
a y F\'\i sica del Espacio (IAFE), UBA-CONICET, CC 67, Suc. 28,
1428 Buenos Aires, Argentina} \affiliation{Departamento de F\'\i
sica, Facultad de Ciencias Exactas y Naturales, Universidad de
Buenos Aires, Argentina}

\begin{abstract}

In this work, we show that when supernova Ia (SN Ia) data sets are
used to put constraints on the free parameters of inhomogeneous
models, certain extra information regarding the light-curve fitter
used in the supernovae Ia luminosity fluxes processing should be
taken into account. We found that the size of the void as well as
other parameters of these models might be suffering extra
degenerations or additional systematic errors due to the fitter. A
recent proposal to relieve the tension between the results from
Planck satellite and SNe Ia is re-analyzed in the framework of
these subjects.

\end{abstract}

\pacs{Valid PACS appear here}

\keywords{Cosmology, inhomogeneous models, Supernovae Ia,
light-curve fitters}  \maketitle

\section{Introduction}

According to a homogeneous and isotropic
Friedmann-Robertson-Walker (FRW) standard model, since 1998
combined observations of nearby and distant type Ia supernovae
(SNe Ia) led to the discovery of the accelerating universe
picture. We now have a \emph{concordance model} in which the
dimming of distant SNe Ia \cite{perlmutter, riess, mik, hicken,
kessler, union2, snls, union21, sdss14, sdss14b}, anisotropies in
the cosmic microwave background (CMB) \cite{wmap9, planck}, and
the signature of baryon acoustic oscillation (BAO) \cite{eisen,
percival} cannot be explained by considering only baryonic and
dark matter. The most popular solution is to introduce an
additional component with negative pressure, the so-called dark
energy (e.g., \cite{ht, sahni, padma, frieman, huterer}).

Some other proposals have been presented since then. Among them,
exact inhomogeneous models with no dark energy component were put
forward shortly after the release of the first supernova
measurements \cite{dabro, pascual, celerier00, tomita}; and more
recently some models, such as the ones based on spherically
symmetric Lema\^{\i}tre-Tolman-Bondi (LTB) and other exact solutions,
began to have an important development in the past few years (see
for example, \cite{celerier7, enqvist, garcia, stephon, biswas,
bolejko1, clarkson, celerier1373, wessel}). Until today, these
have been considered toy-models because of their simplicity and it
is necessary to remark that they are not robust models of our
universe yet. According to some authors (e.g. \cite{gbh, moss}),
LTB models face important observational challenges and the most
simple current versions of these models would be ruled out. As it
has been emphasized in the literature, their use must be
considered as a mere first step towards more sophisticated models
\cite{celerier1373}. Examples somewhat more complex are the ones
known as Swiss-cheese models \cite{brou}, meatball models
\cite{kain} and Szekeres Swiss-cheese models \cite{bolecele}. Note
that LTB model breaks the Copernican principle, by placing the
observer at the centre of a spherically symmetric universe, while
other models, such as the Swiss-cheese one, are only locally
inhomogeneous. For a detailed review of exact solutions, see
\cite{bolejko1}.

The importance of the study of aforementioned models is motivated
by different observational works. In particular, \cite{keenan}
concluded that local measurements of the near-infrared luminosity
density are consistent with models that invoke a large local
under-density (around 300 Mpc) to explain either the apparent
acceleration observed via type Ia supernovae, or to explain the
discrepancy between local measurements of $H_0$ and those inferred
from the CMB. The use of inhomogeneous models is also well
justified even in the standard paradigm. For example, it has
already been demonstrated that although the cosmological
observations are analyzed in the homogeneous framework, matter
inhomogeneities might be mistaken for evolving dark energy
\cite{bolejko2, wessel2}. Other authors have studied how the
presence of a local spherically symmetric inhomogeneity can affect
apparent cosmological observables derived from the luminosity
distance. Under the assumption that the real space-time is exactly
homogeneous and isotropic, they have found that phantom dark
energy or quintessence behaviors can be produced for compensated
underdense or overdense regions \cite{staro}. The fact of putting
observational constraints on these type of models should not be
taken to be ruled out or not, but as a beacon to follow, and to
study possible degenerations present that might influence future
works with more refined proposals.

The possible tension between the best fits for $\Omega_m$ and
$H_0$ obtained from the \emph{Planck} satellite observations on
one hand, and the Hubble diagram of SNe Ia on the other hand, has
been recently faced by the authors in \cite{uzan}. They showed
that the use of an inhomogeneous Swiss-cheese model to interpret
the Hubble diagram allows to reconcile it with \emph{Planck}
results.

The flux measurements from an SN Ia at different epochs and
distinct passbands are processed with the so-called
\emph{light-curve fitters} to obtain luminosity distance values.
The two most used methods are named MLCS2k2 \cite{mlcs} (hereafter
MLCS) and SALT2 \cite{salt2}. Distance moduli calculated for the
same objects by the two fitting methods are not necessarily equal.

MLCS (The Multicolor Light Curve Shape fitter), is the most recent
version of the fitter used by the High-z Supernova Team
\cite{riess}, whilst the SALT2 (Spectral Adaptive Light curve
Template), is an improved version of the fitter used originally by
the Supernova Cosmology Project \cite{perlmutter}. A detailed
description of both fitters and a thorough discussion about
systematic errors in SN surveys can be found for example in
\cite{hicken, kessler}.

It is a known fact that the same SN Ia data set from which
distance estimates are analyzed with two different light-curve
fitters, can lead to different values for various cosmological
parameters, or also some cosmological models would be more favored
than others (e.g. \cite{hicken, kessler, sollerman, sanchez,
carneiro, smale, zhen}). One of us, has recently shown how these
two light-curve fitters employed for the same SN Ia data set
produce the same result than two different SN Ia sets, and it
should be minded as an additional factor to decide whether phantom
type models are favored or not \cite{grb}.

Whereas the MLCS calibration uses a nearby training set of SNe Ia
assuming a close to linear Hubble law, SALT2 uses the whole data
set to calibrate empirical light curve parameters. SNe Ia beyond
the range in which the Hubble law is linear are used, so a
cosmological model must be assumed in the latter. Typically a
$\Lambda$CDM or a $w$CDM ($w=const$) model is assumed.
Consequently, the published values of SN Ia distance moduli
obtained with SALT2 fitter retain a degree of model dependence.
Regarding this issue, in \cite{frieman1} it was pointed out that
systematic errors in the method of SNe Ia distance estimation have
come into sharper focus as a limiting factor in SN cosmology. The
major systematic concerns for supernova distance measurements are
errors in correcting for host-galaxy extinction and uncertainties
in the intrinsic colors of supernovae, luminosity evolution,
U-band rest frame in the low-redshift sample and selection bias.
Also, SALT2 fitter does not provide a cosmology-independent
distance estimate for each supernova, since some parameters in the
calibration process are determined in a simultaneous fit with
cosmological parameters to the Hubble diagram. It is important to
remark that a 0.2 apparent magnitude difference leads to a 10\%
error in the luminosity distance value. Some researchers have
focused on these issues, and important steps have begun to be
taken (e.g. \cite{sullivan, sullivan2, march, lampe, chot, saltmu,
gupta, joha, kessler2, wang, wang2, riga, mosher, sdss14b}).
Unfortunately, these subjects have been boarded, mostly, just from
the SALT2 point of view. Another recently developed light-curve
fitter is SiFTO \cite{sifto}. Although SiFTO differs from SALT2 in
some aspects including improvements with respect to the latter,
SiFTO shares more features with SALT2 than with MLCS (see for
example \cite{snls}). In fact, results from SiFTO are finally
mostly compared using SALT2 as a guide, and the general conclusion
is that the differences associated with these two fitters are not
very significant (e.g \cite{sifto, snls}).

Since in the inhomogeneous framework literature the possible
effects of the mentioned issues have been scarcely studied; and
stimulated by the approach given by \cite{uzan}, in this Letter,
we first analyze the consistency of two of the main light-curve
fitters used for the elaboration of SN Ia data sets in the void
models framework. To accomplish this, we present a study about the
possible results that can be obtained regarding, for instance, the
size of the void; or degenerations in other parameters that are
usually used in this type of models, to parameterize their
profiles or diagnose when viewed as effective models. We also
explore the impact of some systematics induced by the mentioned
fitters in such models, to then study how this could affect
previous proposals in the literature.

Additionally, we propose that the solution found by the authors in
\cite{uzan} might be even more reinforced if the most recent SNe
Ia data were used in the MLCS fitter framework.

\section{The Universe according to void models}

\subsection{Lema\^{\i}tre-Tolman-Bondi model}

Among the variety of papers regarding inhomogeneous models
published in the last few years, we chose to follow
\cite{february} because of the clarity and detail in the presented
results, and because our aim is to show the analysis under
consideration in the framework of a simple model.

Then, following the notation of Section 2 of the mentioned work,
we will describe the observable universe, for this first case
studied, considering an inhomogeneous void centered around us and
adopting a spherically symmetric Lema\^{\i}tre-Tolman-Bondi (LTB) model
which metric is given by
\begin{equation}
ds^2=-dt^2+\frac{a^2_{||}(t,r)}{1-k(r)r^2}\:dr^2+a^2_{\perp}(t,r)r^2\:d\Omega^2
\label{metrica1}
\end{equation}
where the angular ($a_{\perp}$) and the radial ($a_{||}$) scale
factors are related by
\begin{equation}
a_{||}\equiv (a_{\perp}\:r)' \label{factors}
\end{equation}
and a prime denotes partial derivative with respect to coordinate
distance $r$. The curvature $k(r)$ is not constant but is instead
a free function. The coordinates are chosen such that the angular
scale factor is constant and satisfies $a_{\perp}(t_0,r)=1$ at
present epoch. From both scale factors, we can define two Hubble
rates,
\begin{equation}
H_{\perp}=H_{\perp}(t,r)\equiv
\frac{\dot{a}_\perp}{a_\perp},\:\:\:\:\: H_{||}=H_{||}(t,r)\equiv
\frac{\dot{a}_{||}}{a_{||}} \label{hrates}
\end{equation}
where an over-dot indicates partial differentiation with respect
to $t$. When the parameters are evaluated to the time today we
designate them as $H_{\perp0}=H_{\perp0}(r)=H_{\perp}(t_0,r)$ etc.
The Friedmann equation in this geometry is written as
\begin{equation}
H_{\perp}^2=\frac{M}{a_\perp}^3-\frac{k}{a_{\perp}^2},\label{fried}
\end{equation}
where $M(r)$ is another free function of $r$, and the locally
measured energy density is
\begin{equation}
8\pi G \rho(t,r)=\frac{(M r^3)_{,r}}{a_{||} a_{\perp}^2
r^2}\label{energy}
\end{equation}
which satisfies the conservation equation
\begin{equation}
\dot{\rho}+(2 H_\perp+H_{||})\:\rho=0
\label{conser}
\end{equation}

Similarly, as in the case of the FRW models, the dimensionless
density parameters for the curvature and matter are defined as
\begin{eqnarray} \nonumber
&&\Omega_k(r)=-\frac{k}{H_{\perp0}^2}\\
&&\Omega_m(r)=\frac{M}{H_{\perp0}^2} \label{omegas}
\end{eqnarray}
so, the Friedmann equation takes the known form:
\begin{equation}
\frac{H_\perp^2}{H_{\perp0}^2}=\Omega_m a_\perp^{-3}+\Omega_k
a_\perp^{-2}
\label{fried2}
\end{equation}
in such way that $\Omega_m(r)+\Omega_k(r)=1$ is satisfied.
Integrating the Friedmann equation from the big bang time
$t_B=t_B(r)$ to some later time $t$, the age of the universe at a
given ($t,r$) can be obtained by,
\begin{equation}
\tau(t,r)=t-t_B=\frac{1}{H_{\perp0}(r)}\int_0^{a_\perp(t,r)}\frac{dx}{\sqrt{\Omega_m(r)x^{-1}+\Omega_k(r)}}
\label{age}
\end{equation}

We set $t_B=0$ so our model evolves from a perturbed FRW model at
early times. This way, the age of the universe $\tau$ is constant,
and equal to the time today $t_0$. Solving (\ref{age}) for
$H_{\perp0}(r)$, and for the case in which $\Omega_k>0$, we have:
\begin{equation}
H_{\perp0}(r)=\frac{\sqrt{\Omega_k}-\Omega_m\:{\rm
sinh}^{-1}\sqrt{\frac{\Omega_k}{\Omega_m}}}{t_0\:\Omega_k^{3/2}}
\label{hubble}
\end{equation}

We will use the notation for the Hubble constant
$H_0=H_{\perp0}(r=0)$, which fixes $t_0$ in terms of $H_0$,
$\Omega_m(r=0)$ and $\Omega_k(r=0)$.

Following \cite{enqvist,february}, on the past light cone a
central observer may write the ($t,r$) coordinates as functions of
redshift $z$. These functions are determined by the differential
equations,
\begin{eqnarray}
&&\frac{dt}{dz}=-\frac{1}{(1+z) H_{||}}\\
&&\frac{dr}{dz}=\frac{\sqrt{1-k r^2}}{(1+z) a_{||} H_{||}}
\label{ecdif}
\end{eqnarray}
where $H_{||}(t,r)=H_{||}(t(z),r(z))=H_{||}(z)$, etc. The area
distance is given by
\begin{equation}
d_A(z)=a_\perp(t(z),r(z))\:r(z) \label{da}
\end{equation}
and the luminosity distance is $d_L(z)=(1+z)^2\:d_A(z)$. With
these quantities, the distance modulus is given by
\begin{equation}
\mu(z)=m-{\cal{M}}=5\:log_{10}\Big[\frac{d_L(z)}{1\:\text{Mpc}}\Big]+25
\label{mu}
\end{equation}
where $m$ is the apparent magnitude of a source which absolute
magnitude is $\cal{M}$.

For our LTB case analysis, we chose the model 3 of \cite{february}
whose void profile parametrization is given by:
\begin{equation}
\Omega_m(r)=\Omega_{\rm out}-(\Omega_{\rm out}-\Omega_{\rm
in})\:\frac{\sigma^2}{\sigma^2+r^2} \label{void3}
\end{equation}

In the last equation, $\Omega_{\rm in}$ is the value of $\Omega_m$
at the center of the void. As in the cited paper, we fixed
$\Omega_{\rm out}=1$, so that the space is asymptotically flat.
This choice, for us, is yet less relevant than for the authors of
the mentioned work, since here we are not interested in how
realistic is the model or not. The parameter $\sigma$
characterizes the size of the void. Note that $\sigma$ has
dimensions of length (e.g. Mpc). Such profile is capable of
reproducing the $\Lambda$CDM distance modulus to high accuracy.

The selection of model 3 of \cite{february} for our analysis does
not have a specific motivation. We simply choose the model that
the authors find to be the most favored by information criteria
(see Table 6 of \cite{february}). This choice always leads to
$\Omega_k > 0$. As explained in the following section, the
selected model does not have relevance in the goal of this work,
since we are not interested in putting constraints neither finding
the best fits to cosmological models, but to show that certain
extra information should be considered when using SNe Ia data that
has been analyzed with distinct light-curve fitters.

Since there are still only toy models to describe voids,
constructing diagnostics from $\Lambda$CDM allow us to visualize
what our real constraints on $\Lambda$CDM are. Among some
quantities that are usually used as non-concordance diagnostics to
distinguish between FRW/$\Lambda$CDM models and LTB models, we
consider the effective deceleration parameter $q_{\rm eff}(z)$,
and the effective dark energy equation of state for the void model
$w_{\rm eff}(z)$.

These parameters are defined as:
\begin{equation}
q_{\rm eff}(z)=-1+\frac{(1+z)}{H_{||}(z)}\:\frac{d}{dz}H_{||}(z)
\label{qeff}
\end{equation}

\begin{equation}
w_{\rm eff}(z)=\frac{2(1+z)d_c''+3\: d_c'}{3\:[H_0^2\: \Omega_m
(1+z)^3 d_c'^2-1]\:d_c'} \label{weff}
\end{equation}
where in the last equation, $d_c=(1+z)d_A$ is the comoving angular
diameter distance evaluating the void parameters obtained from the
best-fitting model to the data, while $\Omega_m$ and $H_0$
correspond to the best-fitting to the same data, but in the flat
FRW model framework $\Lambda$CDM. A prime here means derivative
with respect to the redshift $z$.

\subsection{Swiss-cheese model reloaded}

Recently, in the work \cite{uzan1}, the authors inferred a
phenomenological expression for the distance-redshift relation in
a Swiss-cheese universe. They found that the luminosity distance
$d_L=(1+z)^2\:d_A^{\:\rm SC}$ can be calculated using the
heuristic linear combination:
\begin{equation}
d_A^{\:\rm SC}(z)=(1-f)\:d_A^{\:\rm holes}(z)+f\:d_A^{\:\rm
FRW}(z) \label{dasc}
\end{equation}
where $d_A^{\:\rm FRW}$ is the angular distance for the FRW case,
$d_A^{\:\rm holes}$ is given by
\begin{equation}
d_A^{\:\rm holes}(z)=\int_0^z\frac{dz'}{(1+z')^2\:H(z')}
\label{daholes}
\end{equation}
and $f$ is the \emph{smoothness parameter} defined by
\begin{equation}
f\equiv \lim_{V\to\infty} \frac{V_{\rm FRW}}{V} \label{fpar}
\end{equation}
with $V_{\rm FRW}$ being the volume occupied by the FRW region
within a volume $V$ of the Swiss-cheese. With this definition,
$f=1$ corresponds to a model with no hole (i.e. a FRW universe)
while $f=0$ corresponds to the case where matter is exclusively
under the form of clumps. Since here we will consider the flat FRW
case with matter ($\Omega_m$) and a cosmological constant
($\Omega_{\Lambda}$), the Hubble parameter is:
$H(z)=H_0\:[\Omega_m(1+z)^3+\Omega_{\Lambda}]^{1/2}$.

In the next section, we will use the LTB model to analyze global
effects and degenerations present when SNe Ia observations
processed with two different light-curve fitters are used. Then,
we will show how the results in the Swiss-cheese model framework
are affected.

\section{Light-curve fitters in void models}

In this section, we will use a $\chi^2$ statistic to analyze the
confidence intervals of the free parameters of the two
cosmological models introduced before, by employing the same SN Ia
data sets, but processed by two different fitters. The three free
parameters are, for the LTB model, $H_0$, $\Omega_{\rm in}$ and
$\sigma$; whilst for the Swiss-cheese model case they are
$\Omega_m$, $H_0$ and $f$. Therefore, these models have the same
number of free parameters as the curved $\Lambda$CDM model cases.

In this work, the analysis was performed in the framework of SALT2
\cite{salt2} and MLCS \cite{mlcs} fitters and the SN Ia data set
used was the SDSSII full data set (Tables 10 and 14 from
\cite{kessler} with the same 'intrinsic' dispersions used there).
This data set is, until today, the best one (publicly available)
treated and analyzed with both fitters.

As already mentioned, here we are not interested in putting
constraints neither in finding the best cosmological model, but to
show certain extra information that should be considered when
using SNe Ia data processed with different fitters.

\begin{figure}[h!]
\begin{center}
\includegraphics[width=9cm,angle=0]{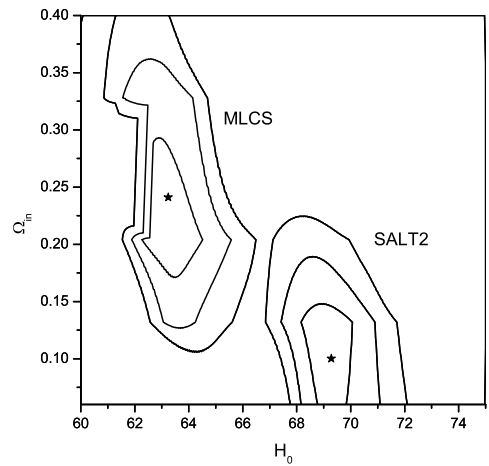}
\end{center}
\caption{ Confidence intervals at 68.3\%, 95.4\% and 99.7\% in the
$H_0-\Omega_{\rm in}$ plane for the SDSSII (MLCS and SALT2) SN Ia
data set in the LTB model framework. The best fits are indicated
with a star. There can be seen a tension between both processing
methods with a confidence level greater than 99.7\%. Values of
$H_0$ are expressed in km/s/Mpc.} \label{Fig1}
\end{figure}

We will start considering the LTB model. In Table \ref{bfits}, the
best-fitting void model parameters derived from SN Ia data are
shown. We can observe that while SALT2 has a tendency to give
lower values of $\Omega_{\rm in}$ and bigger voids, MLCS favors
lower values for $H_0$. In Fig. \ref{Fig1}, we show the confidence
intervals at 68.3\%, 95.4\% and 99.7\% in the $H_0-\Omega_{\rm
in}$ plane for the SDSSII (MLCS and SALT2) SN Ia data set. There
we can see that a tension between the light-curve fitters is
present with more than 99.7\% confidence level. This was one of
the reasons that motivated us to extend our study on inhomogeneous
models to the one presented by \cite{uzan} as we will see further
on.

In \cite{february}, the authors found that when combining $H(z)$
data with SN Ia ones for their analysis, the best fit for the size
of the void of the model considered here, corresponded to a void
380 Mpc bigger than the one obtained with only SNe Ia data. This
behavior was expected, since the fit to $H(z)$ by itself favors
enormous voids. What is interesting is that we found here a
variation of around 350 Mpc by just changing the way of processing
the same SNe Ia data set. Hence, it seems that an additional
uncertainty in $\sigma$ of about 11\% could be associated with the
selection of the fitter used. Therefore, when one seeks to
constraint the typical size of a void, there seems to exist an
extra degeneration between the inclusion of $H(z)$ data and the
fitter employed in the SNe Ia light-curves processing.

\begin{table}[h!]
\caption {Best fits obtained for $H_0$ (km/s/Mpc), $\Omega_{\rm
in}$, and the size of the void $\sigma$ (Mpc) associated with the
LTB model, for the SDSSII (MLCS and SALT2) SNe Ia.}

\centering
\begin{tabular}{|c|c|c|c|} \hline\hline
\:\:\:\bf Fitter \:\:\:& \:\:\:$\bf H_0 $ \:\:\:& \:\:\:$\bf
\Omega_{in}$\:\:\: & \:\:\:$\bf \sigma
$\:\:\:\\
\hline \:\:\:MLCS\:\:\: & \:\:\:63.2 \:\:\: &\:\:\: 0.24\:\:\: &
\:\:\:3050\:\:\:
\\
\:\:\:SALT2 \:\:\:& \:\:\:69.2 \:\:\:& \:\:\:\: 0.10 \:\:\:& \:\:\:3400\:\:\: \\
\hline\hline
\end{tabular}
\label{bfits}
\end{table}

In \cite{february}, the degenerations known between the parameters
of these models are discussed. It is mentioned that to achieve a
similar $\chi^2$, if one wishes to obtain a bigger void (larger
$\sigma$) a lower $\Omega_{\rm in}$ is needed; while emptier voids
(lower $\Omega_{\rm in}$) will require a larger $H_0$. We will see
that the fitters used for the processing of the data lead to
similar degenerations.

A work that considered the dependence of the results on the fitter
employed, in a inhomogeneous model framework, and with the same
SNe Ia set that in the present work, was from the authors of
\cite{sollerman}. In Fig. 5 of that work, there can be seen that
SALT2 favors values of $\Omega_{\rm in}$ lower than those
associated with MLCS. Therefore, according to what was mentioned
earlier, one would expect that SALT2 would prefer larger voids.
And certainly, this is what we found. But we remark that here this
is not a consequence of the way these void models are built, but
it is a matter of data processing.

\begin{figure}[h!]
\begin{center}
\includegraphics[width=8cm,angle=0]{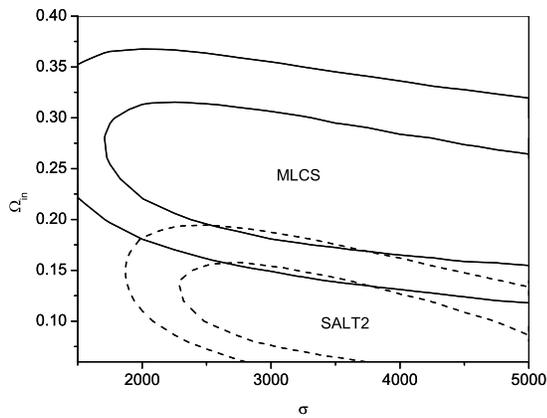}
\end{center}
\caption{Confidence intervals at 68.3\% and 95.4\% in the
$\sigma-\Omega_{\rm in}$ plane for the SDSSII (SALT2) SNe Ia
(dashed lines), and for the SDSSII (MLCS) SNe Ia (solid lines).
The $\sigma$ values are expressed in Mpc.} \label{Fig2}
\end{figure}

We will now analyze the tension between the two light-curve
fitters from another perspective. The mentioned degeneration
between the parameters $\Omega_{\rm in}$ and the size of the void
$\sigma$ can be seen, slightly, in Fig. \ref{Fig2} both for SALT2
and for MLCS.

Let us suppose now that, in the search to find a void for which
$H_0$ is in agreement with the Hubble parameter value found by the
\emph{Planck} Collaboration \cite{planck} or by the nine years of
WMAP \cite{wmap9}, we decided to fix the value of $H_0$ and allow
the variation of the two other free parameters. The reader should
always have in mind that here we are not interested in the best
fits themselves, neither in if the models will be competitive or
not. Also, mind that the data is always the same, and the only
thing that changes and by which the results become altered, is the
way they have been processed. What we would like to address in the
following is, how in the very same situation (fixing $H_0$), the
\emph{same} data behave in a very different way.

In Table \ref{fixho}, we show the best fits to the LTB model
having fixed the value of $H_0$ for the \emph{Planck} and for the
WMAP9 cases. The results are surprisingly different. We know that
there is a degeneration between $H_0$ and the void size $\sigma$.
However, let us note how different it is exhibited when the same
data are changed only by the way of processing.

\begin{table}[h!]
\caption{Best fit values having fixed $H_0$ according to the
\emph{Planck} case \cite{planck} (67.3 km/s/Mpc) or WMAP9
\cite{wmap9} (70.0 km/s/Mpc). The values of $\sigma$ are expressed
in Mpc.}

\centering
\begin{tabular}{|c|c|c|} \hline\hline
 \:&\bf MLCS & \bf SALT2 \\
$H_0 $\:\:\:& $\Omega_{\rm in}$\:\:\:\:\:\:\:\:\:\:$\sigma$ & $\Omega_{\rm in}$\:\:\:\:\:\:\:\:\:$\sigma$\\
\hline \:\:\:67.3 \:\:\:& 0.18\:\:\:\: 1250 & 0.12\:\:\:\:5250 \\
\:\:\:70.0 \:\:\:& 0.16\:\:\:\:\:\:\:750 & 0.10\:\:\:\:2750 \\
\hline\hline
\end{tabular}
\label{fixho}
\end{table}

In Fig. 4 of \cite{february}, the authors show the effect on the
distance modulus $\mu$ for different values of $H_0$. They mention
that the void parameters they obtained are partially dependent on
the value of $H_0$. Here, we obtained that the value of $H_0$ has
higher or lower impact depending on the fitter used. For instance,
for $H_0=67.3$ km/s/Mpc the best fit for $\sigma$ between one
fitter or the other differs by about 420\% and around 370\% for
the value of WMAP9 (see Table \ref{fixho}).

\begin{figure}[h!]
\begin{center}
\includegraphics[width=8cm,angle=0]{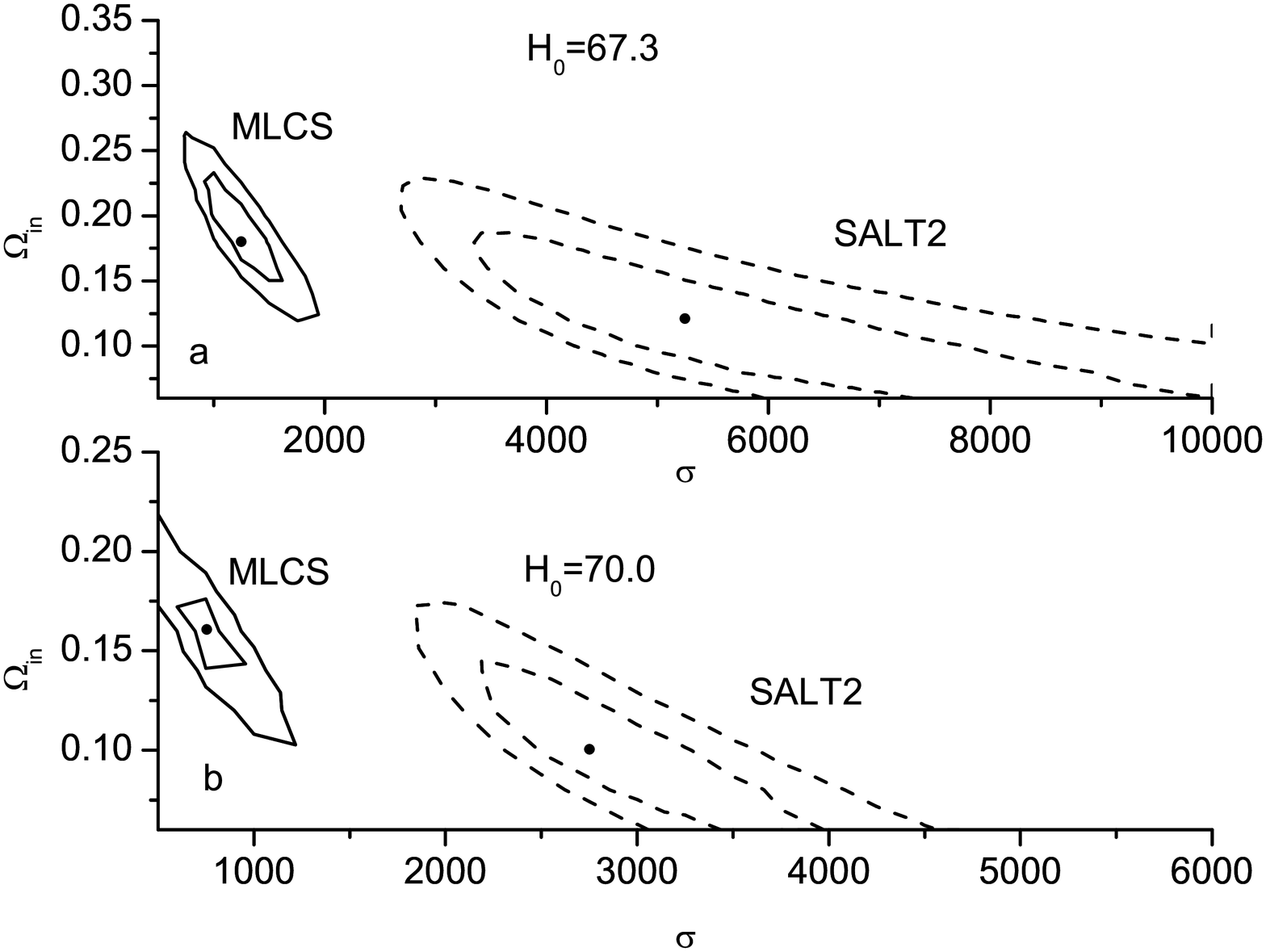}
\end{center}
\caption{(a) Confidence intervals at 68.3\% and 95.4\% in the
$\sigma-\Omega_{\rm in}$ plane for the SDSSII (SALT2) SNe Ia
(dashed lines) and for the SDSSII (MLCS) SNe Ia (solid lines).
This case with $H_0=67.3$ km/s/Mpc. (b) Confidence intervals at
68.3\% and 95.4\% in the $\sigma-\Omega_{\rm in}$ plane for the
SDSSII (SALT2) SNe Ia (dashed lines) and for the SDSSII (MLCS) SNe
Ia (solid lines). This case with $H_0=70.0$ km/s/Mpc. Values of
$\sigma$ are expressed in Mpc, and the best fits are indicated
with a dot.} \label{Fig3}
\end{figure}

In Fig. \ref{Fig3}, we can observe the effect of having fixed
$H_0$. Both for SALT2 and for MLCS, the degeneration between
$\Omega_{\rm in}$ and $\sigma$ appears clearer than in Fig.
\ref{Fig2}. When $H_0$ is not fixed, the ellipses of the
confidence regions are very large. In particular, for the MLCS
case, they are bigger. When fixing the value of $H_0$, as it is
expected, the confidence regions are reduced giving more
restricted values for $\sigma$. However, see how in Fig.
\ref{Fig3} the ellipse corresponding to the MLCS case is the one
which reduces the most, the one which gets the highest impact and
in a different way than SALT2. MLCS seems to constraint the values
of allowed $\sigma$ in a more notorious way than SALT2. It can
also be appreciated how the confidence intervals for the two
fitters in the $\sigma-\Omega_{\rm in}$ plane differ considerably,
indicating the tension between both ways of processing the light
curves of the SNe.

When we fix the value of $H_0$ at 70.0 km/s/Mpc \cite{wmap9}, the
size of the void for the MLCS case is reduced to only 750 Mpc.
What we found here is that MLCS would allow the voids not to be
giant as it is generally suggested in the literature (even though
some authors have shown that giant voids are not mandatory to
explain the observations with a LTB model \cite{nogiant}). This is
not a conclusive assertion, but something we find interesting to
mention as a possible tendency of the data when being processed
with MLCS.

Regarding the value of $H_0$ in the LTB models framework, we would
like to leave a concern raised. In the works \cite{romano}, the
author using a local redshift expansion for the luminosity
distance and a constraint on the age of the universe, showed that
the parameters defining a general LTB model give them enough
freedom to enable them to agree with any value of $H_0$. But if we
manage to suit the value of $H_0$, we wonder: which SNe light
curve processing method should we use to put constraints on the
rest of the free parameters of the model? The fitters might give
very different values to, for example, $\Omega_{\rm in}$ and
$\sigma$.

We now analyze some quantities that are usually used as
non-concordance diagnostics to distinguish between
FRW/$\Lambda$CDM models and LTB models.

\begin{figure}[h!]
\begin{center}
\includegraphics[width=8cm,angle=0]{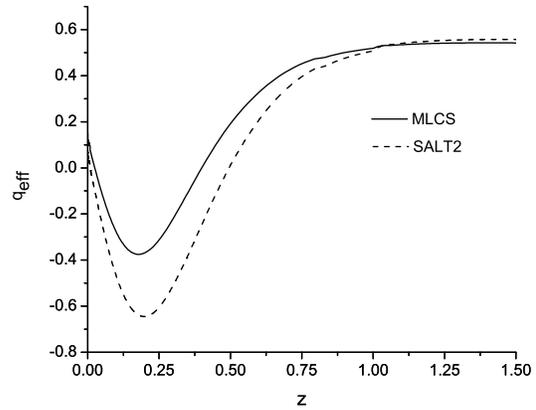}
\end{center}
\caption{Curves of the effective deceleration parameter $q_{\rm
eff}$ vs. $z$ for the best fit parameters of the LTB model
analyzed, to the SNe Ia data processed by SALT2 and MLCS.}
\label{Fig4}
\end{figure}

We did not find significant differences between the $w_{\rm eff}$
vs. $z$ curves obtained with SALT2 and MLCS fitters. Nevertheless,
in the case of the effective deceleration parameter $q_{\rm
eff}(z)$, we did find differences. In Fig. \ref{Fig4}, the $q_{\rm
eff}$ vs. $z$ curves for both light-curve fitters are shown.

In the MLCS case, we see a more pronounced tendency to a
deceleration today than the one found in \cite{february}. Note
that the shape of the curve and the deceleration today found by
the mentioned authors are very similar to the ones found here
under the framework of SALT2 (since both SNe Ia sets are processed
with SALT2). Other authors have already found that when the
supernovae are processed with MLCS, and then combined with other
observations (BAO+CMB+LT) in dark energy models, a case with
deceleration today is favored \cite{zhen}.

As mentioned above, the displayed in Fig. \ref{Fig1} in the
framework of the chosen LTB model, motivated us to analyze the
recent proposal raised in \cite{uzan} to relieve the tension
between the best fits for ($\Omega_m, H_0$) derived from SNe Ia
data and the ones corresponding to the results of the
\emph{Planck} satellite. These authors analyzed the Swiss-cheese
model of \cite{uzan1} with the SNLS 3 data set \cite{snls}
processed with SiFTO/SALT2, and showed that using such
inhomogeneous model to interpret the Hubble diagram allows to
reconcile it with the \emph{Planck} results. Since SNLS 3 data
processed with MLCS are not publicly available, here we will make
the study of the same model, but with SDSSII data \cite{kessler}
processed with both SALT2 and MLCS.

\begin{figure}[h!]
\begin{center}
\includegraphics[width=10cm,angle=0]{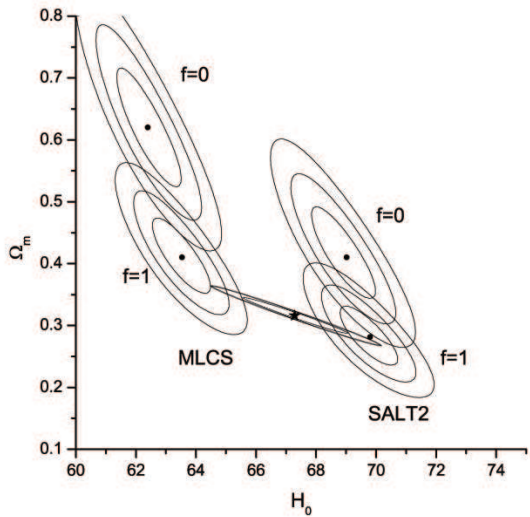}
\end{center}
\caption{Confidence intervals at 68.3\%, 95.4\% and 99.7\% in the
$H_0-\Omega_{m}$ plane for the SDSSII \cite{kessler} (MLCS and
SALT2) SN Ia data set in the Swiss-cheese model framework. Contour
plots with the smoothness parameter $f=1$ correspond to the FRW
case, whilst the ones with $f=0$ correspond to the case where
matter is exclusively under the form of clumps. The best fits are
indicated with a black dot. Values of $H_0$ expressed in km/s/Mpc.
The best fit of Planck is indicated with a star, and the
confidence intervals at 68.3\% and 95.4\% are also shown. No
tension is observed with the data of SDSSII.} \label{Fig5}
\end{figure}

Figure \ref{Fig5} shows the confidence intervals at 68.3\%, 95.4\%
and 99.7\% in the $H_0-\Omega_{m}$ plane for the SDSSII (MLCS and
SALT2) SN Ia data set in the Swiss-cheese model framework of
\cite{uzan1, uzan}.

As we have already mentioned, MLCS favors lower values for $H_0$
(and larger values for $\Omega_m$) therefore the confidence
intervals have been displaced to the left (and up) with respect to
the ones for SALT2. Again, with more than 99\% confidence level,
both light-curve fitters present tension between the obtained
results; and our findings suggest that such fitters might play an
important role in the results or in the conclusions of proposals
such as the ones in \cite{uzan}.

The SNe Ia data of the SDSSII in the framework of a FRW model
(case with $f=1$ in Fig. \ref{Fig5}) do not present tension with
the results of \emph{Planck}, and from this point of view, there
does not seem to be a need for appealing to an inhomogeneous model
to relieve a tension. However, we wonder: could SNLS 3 data
processed with MLCS displace to the left as the ones of SDSSII did
and achieve a better align with \emph{Planck} when taking into
account the suggestion of \cite{uzan} with $0<f<1$? Could a better
compatibility between SNe Ia and CMB be achieved following as a
guide the search of systematics between MLCS and SALT2 and
reducing them? Maybe, the solution found by the authors might be
even more reinforced if the most recent SNe Ia data were used in
the MLCS fitter framework.

Taking into account the results presented in this work, it is
worth making some final reflexions. The community at large uses
public data to put constraints on different proposed models, such
as SNLS \cite{snls}, SDSS \cite{kessler, sdss14}, Union2
\cite{union2}, etc., processed with SALT2. It would be useful and
interesting to have publicly available the same data, in all
cases, also processed with MLCS to be able to use them in carrying
out tests to models alternative to $\Lambda$CDM. As some authors
remarked, the published tables of SNe Ia distance moduli obtained
with the SALT/SALT2 fitters retain a degree of model dependence
(e.g. flat $w$CDM) \cite{smale} and should not be applied to
constrain other models \cite{frieman1}. In recent years, a great
effort has been made to find and study possible sources of
systematic errors \cite{sullivan, sullivan2, march, lampe, chot,
saltmu, gupta, joha, kessler2, wang, wang2, riga, mosher, sifto,
sdss14b}, but most of these works are in the framework of SALT2
and we are still not sure if a light-curve fitter is better than
the other. In \cite{saltmu}, the authors have taken the first
steps towards a possible way to detach of the assumed cosmological
model dependency in the SALT2 processing; and an interesting study
has been made recently with data of the three seasons from the
SDSSII and SNLS, comparing results between MLCS and SALT2
\cite{sdss14b}, but only the distance moduli in SALT2 were
published.

\section{Conclusions}

In the FRW framework where the universe is isotropic and
homogeneous, this is going through an accelerated stage because of
the existence of what we call dark energy. Although the evidence
seems to be solid from various observational data sets, the search
of other alternatives which also explain these observations have
been in development in the last few years. That is the case of the
so-called inhomogeneous models. Even though these are toy-models
yet and do not accomplish the description of the observations
correctly, several authors have remarked the importance of their
study, and even in the standard paradigm, some degree of
inhomogeneity might have a detectable effect on certain observable
quantities.

The luminosity distance measurements of SNe Ia constitute the most
used data sets to put observational constraints on cosmological
models, since they have been and still are the most solid evidence
of the acceleration of the universe detected in the framework of a
FRW model. But as it is well known in the standard paradigm, when
the same SNe Ia data set is processed with two different
light-curve fitters (i.e. SALT2 and MLCS), the values found for
cosmological parameters (such as the equation of state $w$ of dark
energy) differ.

In this work, we analyzed the aforementioned difference showing
that, similarly to what occurs in the standard model, the
light-curve fitters lead to incompatibilities when SNe Ia data are
used to put constraints on inhomogeneous models. This can be seen,
for instance, in the $H_0-\Omega_{\rm in}$ plane in Fig.
\ref{Fig1}.

We found that when the luminosity fluxes coming from supernovae
are processed with the MLCS fitter, the luminosity distances
inferred imply sizes of voids 11\% smaller than in the SALT2 case.
The difference found is of the same order that what other authors
obtained \cite{february} when combining SNe Ia data with data of
$H(z)$. The fitters seem to have a degeneration with these
observations. We also showed that SALT2 favors larger voids and
lower $\Omega_{\rm in}$.

Fig. \ref{Fig1} shows evidence that MLCS favors values of $H_0$
lower than in the SALT2 case (something that also occurs in the
FRW model). This lead us to analyze the proposal of the authors of
\cite{uzan} to relieve the tension in the $H_0-\Omega_{m}$ plane
between the values of the \emph{Planck} Collaboration and the one
from SNLS 3 SNe Ia data \cite{snls}. Although in our work, when
using SDSSII data \cite{kessler}, the mentioned tension with
\emph{Planck} does not appear, maybe the solution found by those
authors might be even more reinforced if the most recent SNe Ia
from SNLS data were used in the MLCS fitter framework. We would
find interesting the release of more public data sets processed
with MLCS to be used in the framework of models alternative to
$\Lambda$CDM. Note that other authors have warned about the risk
of using tables with luminosity distance values in the framework
of SALT2 to put constraints on alternative models (e.g.
\cite{smale, frieman1}). Given that MLCS constitutes a more
model-independent fitter than SALT2, data processed with the
former fitter should be preferentially chosen to put constraints
to alternative models when using SN Ia data. Alternatively, it
would be interesting to have luminosity distances tables in the
framework of proposals like the one developed in \cite{saltmu}.

We found that MLCS tends to favor an effective deceleration today
($q_{\rm eff}>0$) with more emphasis that the SALT2 case. Other
authors have already found these trends, but in the framework of
dark energy models \cite{zhen}.

When analyzing cases in which the value of $H_0$ is fixed (for
example, to see what would happen if one would want to make the
mentioned value compatible with the one obtained from CMB), as it
is expected, the allowed ranges for the size of the void are
reduced. But, in the same situation (fixing $H_0$), both fitters
lead to very different results although the very same data has
been used. Also, for MLCS the ranges associated with the size of
the void get more restricted than in the case of SALT2. It is
interesting to highlight that the size of the void for MLCS, under
this situation, might even be smaller than 1 Gpc; indicating that
when using MLCS, the size of the voids might not need to be so
giant as it is usually sustained.

\bigskip

\acknowledgments{G.R.B. and M.E.D.R. are supported by CONICET
(Argentina). G.R.B. acknowledge support from the PIP
2009-112-200901-00594 of CONICET (Argentina) and M.E.D.R.
acknowledge support from the PIP 2009-112-200901-00305 of CONICET
(Argentina) and the PICT Raices 2011-0959 of ANPCyT (Argentina).
We would like also to thank Sean February for his helpful
clarifications and Diego Travieso for his early collaboration and
interesting discussions.}


\begin{thebibliography}{99}

\bibitem{perlmutter} S. Perlmutter,  et al., Bull. Am. Astron. Soc.
\textbf{29}, (1997) 1351; Astrophys. J. \textbf{517}, (1999) 565;

\bibitem{riess} A. G. Riess, et al., Astron. J. \textbf{116}, (1998) 1009; Astron.
J. \textbf{607} (2004) 665.

\bibitem{mik} G. Miknaitis, et al., Astrophys. J. \textbf{666}, (2007) 674.

\bibitem{hicken} M. Hicken, et al., Astrophys. J. \textbf{700}, (2009) 1097.

\bibitem{kessler} R. Kessler, et al., Astrophys. J. Suppl. Ser. \textbf{185}, (2009) 32.

\bibitem{union2} R. Amanullah,  et al., Astrophys. J. \textbf{716}, (2010) 712.

\bibitem{snls} A. Conley, et al., Astrophys. J. Suppl. \textbf{192}, (2011) 1.

\bibitem{union21} N. Suzuki, et al., Astrophys. J. \textbf{746}, (2012) 85.

\bibitem{sdss14} M. Sako, et al., arXiv:1401.3317, (2014).

\bibitem{sdss14b} M. Betoule, et al., arXiv:1401.4064, (2014).

\bibitem{wmap9} G. Hinshaw, et al., Astrophys. J. Suppl., \textbf{208}, (2013) 19.

\bibitem{planck} Planck Collaboration, paper XVI, arXiv:1303.5076, (2013).

\bibitem{eisen} D. Eisenstein, et al., Astrophys. J. \textbf{633}, (2005) 560.

\bibitem{percival} W. J. Percival, et al., MNRAS \textbf{401}, (2010) 2148.

\bibitem{ht} D. Huterer and M. S. Turner, Phys. Rev. \textbf{D60}, (1999) 081301.

\bibitem{sahni} V. Sahni and A. A. Starobinsky, Int. J. Mod. Phys. \textbf{D9}, (2000) 373.

\bibitem{padma} T. Padmanabhan, Phys. Rep. \textbf{380}, (2003) 235.

\bibitem{frieman} J. Frieman, M. S. Turner and D. Huterer, Annu.
Rev. Astron. Astrophys. \textbf{46}, (2008) 385.

\bibitem{huterer} D. Huterer, \emph{The Accelerating Universe}, arXiv:1010.1162.

\bibitem{dabro} M. P. Dabrowski, Astrophys. J. \textbf{447}, (1995)
43; M. P. Dabrowski and M. A. Hendry, Astrophys. J. \textbf{498},
(1998) 67.

\bibitem{pascual} J. F. Pascual-Sanchez, Mod. Phys. Lett.
\textbf{A14}, (1999) 1539.

\bibitem{celerier00} M.-N. Celerier, Astron. and Astrophys. \textbf{353}, (2000) 63.

\bibitem{tomita} K. Tomita, Astrophys. J. \textbf{529}, (2000)
382011; MNRAS \textbf{326}, (2001) 287.

\bibitem{celerier7} M.-N. Celerier, New Advances in Physics \textbf{1}, (2007) 29.

\bibitem{enqvist} K. Enqvist, Gen. Rel. Grav. \textbf{40}, (2008) 451.

\bibitem{garcia} J. Garcia-Bellido and H. Troels, JCAP \textbf{0804}, (2008) 003.

\bibitem{stephon} S. Alexander, et al., JCAP \textbf{09}, (2009) 25.

\bibitem{biswas} T. Biswas, et al., JCAP \textbf{11}, (2010) 030.

\bibitem{bolejko1} K. Bolejko, et al., Class. Quantum Grav. \textbf{28} (2011)
164002.

\bibitem{clarkson} C. Clarkson, Comptes Rendus Physique, Vol. \textbf{13}, Issue 6, (2012) 682.

\bibitem{celerier1373} M.-N. Celerier, Astron. and Astrophys. \textbf{543}, (2012)
A71.

\bibitem{wessel} W. Valkenburg, et al., MNRAS \textbf{438}, (2014) L6.

\bibitem{gbh} J. Garcia-Bellido and T. Haugb{\o}lle,  JCAP \textbf{09}, (2008) 016.

\bibitem{moss} A. Moss et al., Phys. Rev. \textbf{D83}, (2011)
103515.

\bibitem{brou} N. Brouzakis, et al., JCAP \textbf{0702}, (2007) 013.

\bibitem{kain} K. Kainulainen and V. Marra , Phys. Rev. \textbf{D80}, (2009) 127301.

\bibitem{bolecele} K. Bolejko and M.-N. Celerier, Phys. Rev. \textbf{D82}, (2010) 103510.


\bibitem{keenan} R. C. Keenan et al., Astrophys. J. \textbf{775}, (2013)
62.

\bibitem{bolejko2} K. Bolejko, Astron. and Astrophys. \textbf{525}, (2011) A49.

\bibitem{wessel2} W. Valkenburg, et al., Phys. Dark Univ. \textbf{2}, (2013) 219.

\bibitem{staro} A. E. Romano, et al., European Physical Journal \textbf{C72}, (2012)
2242.

\bibitem{uzan} P. Fleury, et al., Phys. Rev. Lett. \textbf{111}, (2013)
091302.

\bibitem{mlcs} M. M. Phillips, et al., Astrophys. J. \textbf{413}, (1993)
L105; A. G. Riess, et al., Astrophys. J. \textbf{438}, (1995) L17;
S. Jha, et al., Astrophys. J. \textbf{659}, (2007) 122.

\bibitem{salt2} J. Guy, et al., Astron. and Astrophys. \textbf{466}, (2007) 11.

\bibitem{sollerman} J. Sollerman, et al., Astrophys. J. \textbf{703}, (2009) 1374.

\bibitem{sanchez} J. C. Bueno Sanchez, et al., JCAP \textbf{11}, (2009) 029.

\bibitem{carneiro} C. Pigozzo, et al., JCAP \textbf{08}, (2011) 022.

\bibitem{smale} P. R. Smale and D. L. Wiltshire, MNRAS \textbf{413}, (2011) 367.

\bibitem{zhen} L. Zhengxiang, et al., JCAP \textbf{11}, (2010) 031.

\bibitem{grb} G. R. Bengochea, Phys. Lett. \textbf{B696}, (2011) 5; G. R. Bengochea, BAAA, Vol. \textbf{55}, (2013) 431.

\bibitem{frieman1} J. A. Frieman, AIP Conf. Proc. \textbf{1057}, (2008) 87.

\bibitem{sullivan} M. Sullivan, et al., MNRAS \textbf{406}, (2010) 782.

\bibitem{sullivan2} M. Sullivan et al., Astrophys. J. \textbf{737},
(2011) 102.

\bibitem{march} M. C. March, et al., MNRAS \textbf{437}, (2014) 3298.

\bibitem{lampe} H. Lampeitl, et al., Astrophys. J. \textbf{722},
(2010) 566.

\bibitem{chot} N. Chotard, et al., Astron. and Astrophys. \textbf{529}, (2011) L4.

\bibitem{saltmu} J. Marriner, et al., Astrophys. J. \textbf{740}, (2011) 72.

\bibitem{gupta} R. R. Gupta, et al., Astrophys. J. \textbf{740}, (2011)
92; Astrophys. J. \textbf{741}, (2011) 127.

\bibitem{joha} J. Johansson, et al., MNRAS \textbf{435}, (2013) 1680.

\bibitem{kessler2} R. Kessler, et al., Astrophys. J. \textbf{764}, (2013) 48.

\bibitem{wang} X. Wang, et al., Science \textbf{340} issue 6129, (2013) 170.

\bibitem{wang2} S. Wang and Y. Wang, Phys. Rev. \textbf{D 88},
(2013) 043511.

\bibitem{riga} M. Rigault, et al., Astron. and Astrophys. \textbf{560}, (2013) A66.

\bibitem{mosher} J. Mosher, et al., arXiv:1401.4065, (2014).

\bibitem{sifto} A. Conley et al. Astrophys. J. \textbf{681}, (2008) 482.

\bibitem{february} S. February, et al., MNRAS \textbf{405}, (2010) 2231.

\bibitem{uzan1} P. Fleury, et al., Phys. Rev. \textbf{D87}, (2013) 123526.

\bibitem{nogiant} M.-N. Celerier, et al., Astron. and Astrophys. \textbf{518}, (2010) A21.

\bibitem{romano} A. E. Romano, arXiv:1105.1864, (2011); A.
E. Romano, arXiv:1112.1777, (2011).




\end{thebibliography}
\end{document}